\documentstyle[aps,preprint,epsfig]{revtex}

\begin{document}

\title{{\it Ab Initio} Simulation of the Nodal Surfaces of Heisenberg 
  Antiferromagnets}

\author{R.F. Bishop$^1$, D.J.J. Farnell$^2$, and Chen Zeng$^3$}

\address{$^1$Department of Physics, University of Manchester Institute of
  Science and Technology (UMIST), P O Box 88, Manchester M60 1QD, United 
  Kingdom.}

\address{$^2$Institut f\"ur Theoretische Physik, Universit\"at zu K\"oln, 
  Z\"ulpicher Str., 50937 K\"oln, Germany.}

\address{$^3$Department of Physics and Astronomy, Rutgers University, 
  Piscataway, NJ 08855, United States of America.}

\date{\today}

\maketitle

\abstract
{The spin-half Heisenberg antiferromagnet (HAF) on the square 
  and triangular lattices is studied using the coupled 
  cluster method (CCM) technique of quantum many-body theory. 
  The phase relations between different expansion coefficients 
  of the ground-state wave function in an Ising basis for the 
  square lattice HAF is exactly known via the Marshall-Peierls sign 
  rule, although no equivalent sign rule has yet been obtained 
  for the triangular lattice HAF.  Here the CCM is used to
  give accurate estimates for the Ising-expansion coefficients 
  for these systems, and CCM results are noted to be fully consistent 
  with the Marshall-Peierls sign rule for the square lattice case. 
  For the triangular lattice HAF, a heuristic rule is presented 
  which fits our CCM results for the Ising-expansion coefficients 
  of states which correspond to two-body excitations with respect 
  to the reference state. It is also seen that Ising-expansion 
  coefficients which describe localised, $m$-body excitations 
  with respect to the reference state are found to be highly 
  converged, and from this result we infer that the nodal
  surface of the triangular lattice HAF is being accurately
  modeled. Using these results, we are able to make suggestions 
  regarding possible extensions of existing quantum Monte Carlo 
  simulations for the triangular lattice HAF.}

\pagebreak

\section{Introduction}

The coupled-cluster method\cite{ref1,ref2,ref3,ref4,ref5,ref6,ref7,ref8,ref9}
(CCM) has been previously applied\cite{Roger,Bishop1,Bishop3,Bishop2,Lo1,Bishop4,Farnell1,Xian1,Bursill,Xian,Zeng3,Bishop5,Farnell2,Zeng1,Zeng2,Bishop6} 
to a number of unfrustrated and frustrated lattice quantum spin 
systems with a great deal of success. Recently, it has been shown\cite{Zeng2} 
for a CCM calculation, within the so-called SUB2 approximation 
which contains {\it all} two-body correlations, that the ket-state 
correlation coefficients can be related to the Marshall-Peierls 
sign rule\cite{Marshall1}  for the square lattice Heisenberg 
antiferromagnet (HAF). Furthermore, it was shown in this 
treatment\cite{Zeng2} that the CCM ket-state correlation 
coefficients for the Heisenberg model on the triangular lattice
also demonstrate a seemingly regular pattern. This relationship
between the Marshall-Peierls sign rule and the CCM ket-state
correlation coefficients was further investigated\cite{Bishop6} 
and clarified for a spin model with both nearest-neighbour and 
next-nearest-neighbour exchange, namely the $J_1$--$J_2$ model, 
on the linear chain and square lattices. The Marshall-Peierls
sign rule at the Heisenberg point on the square lattice
was found to be preserved by {\it all} of the Ising-expansion 
coefficients within a localised approximation scheme, as
well as within the SUB2 scheme. No assumption was made other than 
the choice of an initial reference state with respect to which the 
ground-state wave function was approximately constructed 
in the infinite lattice limit using the CCM formalism.
It was also possible to obtain a quantitative value for the 
point at which the sign-rule breaks down with increasing 
antiferromagnetic next-nearest-neighbour exchange 
for the square lattice case. This quantitative value
was found to be in good agreement with exact diagonalisation
calculations\cite{Richter1}. 

The Marshall-Peierls sign rule for Heisenberg antiferromagnets on 
bipartite lattices is an exact statement for the phase relations 
between the expansion coefficients of the ground-state wave function 
in an Ising basis. This statement is of intrinsic interest 
because it provides exact information regarding the ground-state
wave function in cases (e.g., the square lattice HAF) for which 
no general exact solution has yet been determined. It is also
of interest because by using this rule one 
implicitly has full knowledge of the `nodal surface' of the 
ground-state wave function within this basis. 
(Note that whenever the words `nodal surface' are 
referred to in this article, this is taken to mean
the magnitudes and signs of these expansion coefficients.) 
Of course, for the square lattice HAF the signs
are exactly known via the Marshall-Peierls sign rule,
although the magnitudes are known only via exact or 
approximate calculation. Quantum Monte Carlo (QMC) 
calculations for fermionic and spin systems 
demonstrate the infamous ``sign-problem'', and so 
such information regarding the signs of the 
expansion coefficients is of importance in these 
calculations. 

For non-bipartite lattices (e.g., the 
triangular lattice HAF) or for other models containing 
frustration (e.g., the $J_1$--$J_2$ model on the square 
lattice) corresponding exact sign-rule theorems are
generally not known. Hence, various approximate ways 
of simulating the nodal surface have been developed for
use, for example, in fixed-node quantum Monte Carlo
calculations. However, one is always limited in
such simulations by not knowing how close is the 
trial wave function or guiding wave function 
that one uses to the true ground-state wave function.
It is therefore of considerable interest to develop theories which 
simulate the nodal surface from an {\it ab initio} standpoint,
such as is done here via the CCM. In this way, one might 
be able to utilise this CCM information concerning 
the nodal surface in a QMC calculation, or at least
suggest possible extensions of previous QMC calculations.
It is also conceivable that one might observe patterns in the 
expansion coefficients obtained via the CCM and so infer 
rules concerning these coefficients or sub-classes
of them which relate to specific types of excitation.

In this article, we briefly introduce the CCM formalism for
the lattice quantum spin systems and also make an explicit
link between the expansion coefficients of the ground-state
wave function and the CCM ket-state correlation coefficients.
This link is investigated for the square and triangle
lattice HAFs, and the Ising-expansion coefficients 
corresponding to states which correspond to two-body excitations 
with respect to a reference state are determined 
within well-defined approximation schemes. These coefficients 
are seen to be well-converged using a {\it localised} 
approximation scheme. For the triangular lattice HAF, a 
heuristic rule is presented which fits our CCM results for 
the Ising-expansion coefficients of states which correspond 
to two-body excitations with respect to the reference state. 
The Ising-expansion coefficients of states which correspond 
to $m$-body excitations are also shown, within this same 
localised approximation scheme, to be well-converged. 
Patterns are also seen in the Ising-expansion coefficients which 
correspond to these $m$-body excitations for the triangular 
lattice HAF, although no generic sign rule for them is formulated. 
Finally, suggestions are made regarding possible extensions 
of existing quantum Monte Carlo simulations for the 
triangular lattice HAF.

\section{The Spin Model And The CCM Formalism}

\subsection{The Heisenberg Model}

In this paper we consider the zero-temperature ($T=0$) 
properties of the spin-half Heisenberg antiferromagnet 
(HAF) quantum spin system, which is described by the 
Hamiltonian
\begin{equation} 
H=\sum_{\langle i,j \rangle} {\bf s}_i . {\bf s}_j ~~ ,
\label{eq1}
\end{equation}
where the index $i$ runs over all $N$ lattice sites 
on the square and triangular lattices with periodic boundary 
conditions, and the index $j$ runs over all nearest-neighbour 
sites to $i$. The angular brackets indicate that each 
nearest-neighbour bond (or link) is counted once and once only.

The Heisenberg model on the square lattice has not been
solved exactly up till now, although it been extensively studied
using various approximate methods\cite{Anderson1,Singh2,Carlson,Runge1}.
Runge\cite{Runge1} has performed the most accurate Monte
Carlo simulation to date for the square-lattice, isotropic
HAF. He finds a value for the ground-state energy per spin of
$E_g/N$=$-$0.66934(4), and a value for the sublattice magnetisation
which is 61.5$\%$$\pm$0.5$\%$ of the classical value. Extensive 
CCM calculations have also been carried out for this 
model\cite{Zeng1} giving a value for the ground-state energy per spin of
$E_g/N$=$-$0.66968, and a value for the sublattice magnetisation
which is 62$\%$ of the classical value. In
comparison, linear spin-wave theory (LSWT) \cite{Anderson1}
gives a value of $E_g/N$=$-$0.658 for the ground-state energy, and a 
value for the sublattice magnetisation which is 60.6$\%$ of 
the classical value.


Again, no exact solution exists for the Heisenberg model on the 
triangular lattice, although many approximate
calculations have been carried 
out\cite{Zeng1,Huse,Singh1,Jolicoeur1,Boninsegni1,Bernu1,Henley1}. 
Exact series expansion\cite{Singh1} calculations obtain a value for
the ground-state energy per spin of this model of $E_g/N$=$-0.551$.
Similarly, spin-wave theory (SWT)\cite{Jolicoeur1} gives a value
of $E_g/N$=$-0.5388$, and a fixed-node quantum Monte Carlo 
(FNQMC)\cite{Boninsegni1} calculation gives a value of 
$E_g/N$=$-0.5431\pm0.0001$. Exact diagonalisations of finite-sized 
clusters of spins\cite{Bernu1} which have been extrapolated to the 
infinite lattice limit give a value for the ground-state energy 
per spin of $E_g/N$=$-0.5445$. 
In comparison, a recent coupled cluster method (CCM) 
calculation\cite{Zeng1} predicts a 
value for the energy of $E_g/N$=$-0.5505$, which is fully 
consistent which the best of the other results obtained for this 
system so far.

Classically, this system orders as a N\'eel state on the 
triangular lattice such that
each nearest-neighbour pair of spins makes an angle 
of 120$^\circ$ to each other. The results of approximate theories
for the amount of this ordering that remains in the 
quantum limit, i.e., the sublattice magnetisation, are 
typically\cite{Zeng1,Jolicoeur1,Bernu1} that 
50$\%$ of the classical ordering remains in the quantum 
limit. The FNQMC calculation of Ref.\cite{Boninsegni1} 
places this value at 60$\%$, which is probably still too high. 
A notable exception to all of these results quoted above
is the result of exact series expansion calculations\cite{Singh1} 
which predicts that as little as 20$\%$ of the classical 
ordering remains.


\subsection{The CCM Formalism}

The CCM formalism\cite{ref1,ref2,ref3,ref4,ref5,ref6,ref7,ref8,ref9}
is now briefly considered, although the reader should
note that more detailed descriptions of the CCM applied to spin systems 
are given in Refs.\cite{Roger,Bishop1,Bishop3,Bishop2,Lo1,Bishop4,Farnell1,Xian1,Bursill,Xian,Zeng3,Bishop5,Farnell2,Zeng1,Zeng2,Bishop6}. 
To calculate the ground state 
wave function $|\Psi\rangle$ of a spin system we start with a 
model state $|\Phi\rangle$ and a correlation operator $S$ such that
\begin{equation} 
|\Psi\rangle = e^S|\Phi\rangle ~~ ,
\label{eq2}
\end{equation} 
where the ket-state correlation operator $S$ may be written as,
\begin{equation} 
S=\sum_{I\neq 0} {\cal S}_I C^+_I ~~ . 
\label{eq3}
\end{equation}
The correlation operator $S$ is formed from a linear 
combination of multiconfigurational creation operators, 
$\{ C^+_I\}$, (which themselves are formed from products 
of spin raising operators) multiplied with the 
relevant ket-state correlation coefficients $\{ {\cal S}_I \}$. 
In addition, we define a set of destruction operators, 
$\{ C^-_I\}$, which are the Hermitian adjoints of 
$\{ C^+_I\}$. 

For the square lattice case, the model state is chosen
to be the classical N\'eel state. The lattice is divided
into two sublattices and one sublattice is populated
with `up' spins and the other with `down' spins. In order
to treat the spins equivalently the local spin axes 
of the `up' spins are rotated by 180$^{\circ}$ about
the $y$-axis, which is mathematically written as,
\begin{equation} 
s^x \rightarrow -s^x  \;\;  ; \;\; s^y \rightarrow s^y \;\; ; \;\;
s^z \rightarrow -s^z \;\; . 
\label{eqnew1}
\end{equation} 
The Hamiltonian may now be written, with the introduction
of an anisotropy coefficient $x$ on the off-diagonal elements 
of $H$, as
\begin{equation} 
H = - \sum_{\langle i,j \rangle} \biggl \{ s_i^z s_{i+\rho}^z +   
\frac x2 (s_i^+ s_{i+\rho}^+ + s_i^- s_{i+\rho}^-) \biggr \} ~~ ,
\label{eqnew2}
\end{equation} 
where $x=1$ corresponds to the isotropic Heisenberg Hamiltonian 
of Eq. (\ref{eq1}). The Marshall-Peierls sign rule for the original 
Hamiltonian of Eq. (\ref{eq1}), in terms of rotated spin coordinates,  
now differs from the original sign rule\cite{Marshall1}. 
For the ground-state wave function, $|\Psi\rangle = \sum_I 
\Psi_I |I\rangle$, the Ising-expansion coefficients $\{\Psi_I\}$
must now all be greater than (or equal to) zero with respect to a complete
set of Ising basis states $\{|I\rangle\}$ in the rotated 
spin coordinates.

For the triangular lattice case, we choose $|\Phi\rangle$ to again be 
the classical N\'eel state, and we start the CCM calculation 
by dividing the lattice into three sublattices, denoted 
$\{$A,B,C$\}$. The spins on sublattice A 
are oriented along the negative {\em z}-axis, and spins on sublattices 
B and C are oriented at $+120^\circ$ and $-120^\circ$, respectively, 
with respect to the spins on sublattice A. In order both to facilitate 
the extension of the isotropic HAF to include 
an Ising-like anisotropy first introduced by Singh and Huse\cite{Singh1}  
and to make a suitable choice of the CCM model state, we perform 
the following spin-rotation transformations. 
Specifically, we leave the spin axes on sublattice A unchanged, and we 
rotate about the $y$-axis the spin axes on sublattices B and C by 
$-120^\circ$ and $+120^\circ$ respectively, 
\begin{eqnarray} 
s_B^x \rightarrow -\frac{1}{2} s_B^x - \frac{\sqrt{3}}{2} s_B^z \;\;  &;& \;\; 
s_C^x \rightarrow -\frac{1}{2} s_C^x + \frac{\sqrt{3}}{2} s_C^z \;\; , 
\nonumber \\
s_B^y \rightarrow s_B^y \;\; &;& \;\; s_C^y \rightarrow s_C^y \;\; , 
\nonumber \\ 
s_B^z \rightarrow  \frac{\sqrt{3}}{2} s_B^x -\frac{1}{2} s_B^z \;\; &;& \;\; 
s_C^z \rightarrow -\frac{\sqrt{3}}{2} s_C^x -\frac{1}{2} s_C^z \;\; . 
\label{eq4}
\end{eqnarray} 
We may rewrite Eq. (\ref{eq1}) in terms of spins defined in these 
local quantisation 
axes for the triangular lattice with a further introduction of an 
anisotropy parameter $\lambda$ for the non-Ising-like pieces, 
\begin{eqnarray} 
H = \sum_{\langle i\rightarrow j\rangle}
\Bigl\{ 
&& 
-{1\over 2} s_i^z s_j^z
+\frac{\sqrt{3}\lambda}{4}
\left( s_i^z s_j^+ +s_i^z s_j^- -s_i^+ 
s_j^z- s_i^-s_j^z \right) \nonumber \\  
&& 
+\frac{\lambda}{8}
\left( s_i^+s_j^- + s_i^- s_j^+ \right) 
-\frac{3\lambda}{8}
\left( s_i^+ s_j^+ + s_i^- s_j^- \right) 
\Bigl\}  
\;\; ,   
\label{eq5}
\end{eqnarray}
where $\lambda=1$ corresponds to the isotropic Heisenberg Hamiltonian 
of Eq. (\ref{eq1}). We note that the summation in Eq. (\ref{eq5}) 
again runs over nearest-neighbour
bonds, but now also with a {\it directionality} indicated by 
$\langle i \rightarrow j\rangle$, which goes from A to B, B to C, and 
C to A. We note that no exact sign rule has yet been proven
for the Hamiltonian of Eq. (\ref{eq5}).
When $\lambda=0$, the Hamiltonian in Eq. (\ref{eq5}) describes 
the usual 
classical Ising system with a unique ground state which is simply 
the fully aligned (``ferromagnetic'') configuration in the local 
spin coordinates described above. 



From the Schr\"odinger equation, $H|\Psi\rangle=E|\Psi\rangle$, we obtain 
an expression for the ground state energy which is given by
\begin{equation}
E=\langle\Phi| e^{-S}He^S|\Phi\rangle ~~.
\label{eq6}
\end{equation}
This equation illustrates one of the key aspects of CCM, which is
namely the similarity transform -- here of the Hamiltonian. 
The similarity transform of any operator may be expanded as 
a series of nested commutators  such that an exact 
expression for the ground-state energy in terms of the 
ket-state correlation coefficients, $\{ {\cal S}_I \}$, which is
already determined in the infinite lattice limit $N \rightarrow \infty$, 
can often be found for whatever (non-trivial) approximations are 
made for $S$. The approximation in $S$ means that the 
we obtain approximate values for ket-state coefficients,
$\{ {\cal S}_I \}$, and so the ground-state energy is 
approximately determined, although, as stated above, the 
equation for the ground-state energy is an exact
expression of one or more of these coefficients.
To determine the CCM ket-state coefficients we operate 
on the Schr\"odinger equation with $\exp(-S)$ and then by 
$\langle\Phi| C_I^-$, for a given cluster configuration
indicated by the index $I$. 
An explanation of how the CCM equations may be derived
and then solved is presented in Refs. \cite{Roger,Bishop1,Bishop3,Bishop2,Lo1,Bishop4,Farnell1,Xian1,Bursill,Xian,Zeng3,Bishop5,Farnell2,Zeng1,Zeng2,Bishop6}.

The three most commonly employed approximation 
schemes are: (1) the SUB$n$ scheme, in which all correlations involving 
only $n$ or fewer spins are retained, but no further restriction 
is made concerning their spatial separation on the lattice;
(2) the SUB$n$-$m$  sub-approximation, in which all SUB$n$ 
correlations spanning a range of no more than $m$ adjacent lattice sites
are retained; and (3) the localised LSUB$m$ scheme, which retains all
multi-spin correlations over distinct locales on the lattice
defined by $m$ or fewer contiguous sites. Note that all 
$30$ cluster configurations which are used in the LSUB4 
approximation for the triangular lattice antiferromagnet, 
and which are independent under the six-point symmetry 
group inherent in the Hamiltonian of Eq. (\ref{eq5}), are listed in 
Fig. \ref{fig3}.


We now consider an expansion of the ground-state wave function in 
a complete Ising basis $\{ |I\rangle \}$ 
(in terms of the {\it local} coordinates 
after rotation). This may be again written as, $|\Psi\rangle = \sum_I 
\Psi_I |I\rangle$, where the sums over $I$ goes over all $2^N$ Ising
states, and we find that this expression naturally leads from
Eq. (\ref{eq2}) (also see Ref. \cite{Bishop6}) to an exact mapping of the 
CCM correlation coefficients $\{ {\cal S}_I \}$ to the Ising-expansion
coefficients $\{ \Psi_I \}$, which is  given by
\begin{equation}
\Psi_I = \langle \Phi | C_I^- e^S | \Phi \rangle ~ \equiv ~
\langle \Phi | s_{i_1}^- s_{i_2}^- \cdot\cdot\cdot s_{i_l}^- ~
e^S | \Phi \rangle ~ .
\label{eq7}
\end{equation}
It is possible to match the terms in the exponential 
to the `target' configuration of $C_I^-$ in Eq. (\ref{eq7}), and so
obtain a numerical value for the $\{ \Psi_I \}$ coefficients
once the CCM ket-state equations have been derived and solved
for a given value of the anisotropy. As the target configuration is formed
from a finite number of spin lowering operators, there is a cut-off
point in the Taylors series expansion of the exponential 
above which no contribution to $\Psi_I$ in Eq. (\ref{eq7}) 
can possibly occur. This matching may be 
achieved\cite{Bishop6} in two ways: analytically at the 
SUB2 level of approximation or at small LSUB$m$ 
levels of approximation; or by using 
computational-algebraic techniques.
We note that we use, for the sake of consistency, 
the same configurations for the `target' 
configurations in $C_I^-$ in Eq. (\ref{eq7}) 
as are used in the ket-state correlation operator
$S$ for the approximation schemes 
define above. We also note that, although this is an 
approximate calculation, we are already dealing
with the infinite lattice limit, $N \rightarrow \infty$.

The next section describes our results for the set 
of $\{ \Psi_I \}$ coefficients for the spin-half 
Heisenberg model on the square and triangular lattices using the
CCM formalism, and discusses possible rules that
can be inferred from these results.

\section{Results}

\subsection{The Two-body Excitations}

The results for the Ising-expansion coefficients $\{ \Psi_I \}$
corresponding to two-body excitations with respect to the model
state for the spin-half square lattice HAF ($x=1$) are 
shown in Fig. \ref{fig2}. (Note that the solution to the 
CCM equations is tracked from $x=0$, at which point we know
that all of the ket-state correlation coefficients 
$\{ {\cal S}_I \}$ are zero, to the isotropic Heisenberg point.)
The results for the two-body $\{ \Psi_I \}$ coefficients
are then determined and it is found that these coefficients
are {\it all} positive for the LSUB$m$ approximation scheme 
(with $m \leq 8$) and within the SUB2 approximation.
(This result for the SUB2 case was previously seen
in Ref. \cite{Bishop6}.) This is an expression
of the Marshall-Peierls sign rule\cite{Marshall1}
which provides an exact relation between the 
$\{ \Psi_I \}$ expansion coefficients. However,
we note that the CCM predicts such behaviour with 
no recourse to exact proofs, and this prediction
is a natural result of the CCM calculations. 
Furthermore, the convergence of the LSUB$m$ series
of results compared to SUB2 calculation 
can clearly be seen in Fig. \ref{fig2}. 
For short lattice distance, we believe that
the LSUB$m$ results are much better model of the
`nodal surface' of states corresponding 
to two-body excitations than those of the SUB2 
calculation.

For the spin-half triangular lattice HAF ($\lambda=1$),
the results for the Ising-expansion coefficients $\{ \Psi_I \}$
corresponding to two-body excitations with respect to the model
are shown in Fig. \ref{fig1} and in Table 1. 
(Note again that the solution to the 
CCM equations is tracked from $\lambda=0$, at which point again 
we know that all of the ket-state correlation coefficients 
$\{ {\cal S}_I \}$ are zero, to the isotropic Heisenberg point.)
A rule that fits the behaviour seen in Fig. \ref{fig1} can
now be stated as: the coefficients are found to be positive 
if the lattice vector of between the two spin-raising operators
in $S$ connects sites on different sublattices; conversely, 
the $\{ \Psi_I \}$ coefficients are found to be negative if 
the lattice vector connects sites on the same sublattice.
It is again seen that the LSUB$m$ series is clearly 
well-converged for small lattice distance $R$. This series of
LSUB$m$ results also clearly present a much better 
representation of the `nodal surface' for the two-body 
excitations, for small lattice distance than the 
corresponding $\{ \Psi_I \}$ coefficients within the
SUB2 approximation\cite{Zeng2}. 

\subsection{The $m$-body Excitations}

For the Ising-expansion coefficients $\{ \Psi_I \}$
corresponding to $m$-body excitations for the square lattice HAF
($x=1$) , 
it has previously been noticed \cite{Bishop6} that the CCM calculation 
outlined above predicts that they are {\it all} positive. 
This is an indication that the CCM results are (once more) fully
consistent with the exact Marshall-Peierls sign rule for this
model. It should be again noted that the only assumption
made is the initial model state, and that the rest follows
naturally on from the CCM calculation.

Table 1 indicates the $\{ \Psi_I \}$ expansion coefficients for 
for the triangular lattice HAF ($\lambda=1$) 
states corresponding to localised, $m$-body (with $m \leq 4$) 
configurations (i.e., those also used within the LSUB4 approximation) 
with respect to the model state. One should note however 
that these $\{ \Psi_I \}$ expansion coefficients are determined 
within the SUB2 and LSUB$m$ (with $m=\{3,4,5,6\}$)
approximations. It is again seen in Table 1 that all of 
the expansion coefficients are well converged 
for the LSUB5 and LSUB6 levels of approximation. Note, however,
that within the LSUB5 and LSUB6 approximation many
other $\{ \Psi_I \}$ coefficients have been determined,
within the `consistency' assumption explained above.

Patterns in the $\{ \Psi_I \}$ expansion coefficients 
for the triangular lattice HAF ($\lambda=1$) for states 
corresponding to $m$-body excitations (with $m \geq 1$) 
have also been observed. It was seen that, with $m$ odd, all 
of the coefficients for excitation clusters which 
are equivalent under both the symmetries of the 
lattice (a point group of order twelve) and the Hamiltonian 
(a point group of order six), were found to have both 
CCM correlation coefficients and Ising-expansion coefficients 
which are zero at all levels of approximation; e.g., 
configurations (1), (3), and (5) in Fig. \ref{fig3}
and in Table 1. In contrast, those  $\{ \Psi_I \}$
coefficients corresponding to configurations with {\it odd} numbers 
of spins which are equivalent under the symmetries of the 
lattice but are {\it not} equivalent under the symmetries of
the Hamiltonian were found to have equal magnitudes but opposite signs.
That is, there exists pairs ($12/6=2$) of coefficients, for
the odd-number spin-excitation clusters, which differ with 
respect to each other only by a phase factor of $-1$;
e.g., those pairs (7,12), (8,10), (16,19), (18,25),
(21,23), and (27,29) in Fig. \ref{fig3} and in Table 1.
In contrast, for the Ising-expansion coefficients 
$\{ \Psi_I \}$ corresponding to $m$-body excitations, 
with $m$ even, we find that those  $\{ \Psi_I \}$ coefficients 
corresponding to configurations which are equivalent under
symmetries of the lattice but are {\it not} equivalent under the
symmetries of the Hamiltonian are exactly the same. 
That is, there exists pairs ($12/6=2$) of coefficients, for
the even-number spin-excitation clusters, which have 
a phase factor with respect to each other of unity;
e.g., those pairs (9,13), (22,26), and (28,30) in 
Fig. \ref{fig3} and in Table 1.

We note that these simple relations were obtained
with no assumption other than the model state, 
which is of course the classical ground state 
of this system. This is in contrast to variational
calculations\cite{Huse}, spin wave theory\cite{Henley1} 
and FNQMC\cite{Boninsegni1} which assume that
the classical phase relations are correct phase
relations for their quantum mechanical counterparts.
It has been shown \cite{Xian} that in a case in 
which the Hamiltonian shares some symmetry
of the model state then the values of the 
CCM correlation coefficients in the ground state, 
for configurations which are equivalent under these 
symmetries, must be completely identical. 
We believe that the `rules' presented above
for the $m$-body expansion coefficients
for the triangular lattice HAF are therefore
a reflection of the symmetries inherent in the 
Hamiltonian and lattice.
Furthermore, it remains an open question as to the 
solution of {\it all} branches of coupled, non-linear, 
ket-state CCM equations for high-order calculations. 
One should note that it might be possible to obtain 
another solution branch which might {\it not} demonstrate 
the full symmetries of the lattice and Hamiltonian, 
as has been seen for other systems treated by CCM before. 
A full treatment of this subject would 
form the contents of another article.

Apart from these simple observations, we cannot infer a more
general rule as to the behaviour of the expansion coefficients 
for $m$-body excitations with $m > 2$. However, we do
have results for the signs of all of the expansion
coefficients within the LSUB$m$ approximation scheme,
as is illustrated in Table 1, 
and hence the CCM is simulating, from an {\it a priori} 
view-point, the nodal surface of this model. From the 
amount of convergence of the results of the expansion 
coefficients within the LSUB$m$ scheme described above, 
we therefore also believe that we have an accurate 
simulation of the nodal surface of this model using the LSUB$m$ 
approximation scheme.

\subsection{Relevance to QMC Calculations}

For the triangular lattice HAF, a `naive' picture of 
how important each excitation configuration
can be found by looking at the magnitude of its corresponding
$\Psi_I$ coefficient, illustrated in Table 1.
In this manner, we find that the two-body,
nearest-neighbour configuration (configuration (2) in 
Fig. \ref{fig3}) is the most important correlation
for this model. By this rationale, the next most important
configurations are a `diamond-shaped', four-body 
configuration (configuration (11) in Fig. \ref{fig3}) and 
a `dog-leg', three-body configuration (configurations (8) 
and (10) in Fig. \ref{fig3}). 
A previous FNQMC calculation\cite{Boninsegni1} contains
all two-body correlations and this particular three-body correlation,
and so a straightforward extension of this calculation 
would be to include this four-body configuration. However, from this
`naive' interpretation of our results it appears that, as well as 
the two-body configurations and the two configurations
mentioned above, many other {\it localised} three-body and 
four-body configurations are also very important.
(These configurations are localised in the same sense of those
configurations contained in the LSUB$m$ approximation.)
Hence, to attempt an accurate simulation of the this model one 
probably needs to find a way of also including many, if not all, 
of these localised, higher-order correlations. One way of achieving this
would therefore be to use the nodal surface predicted by the CCM 
in a FNQMC calculation.

\section{Conclusions}

In this article it has been shown that the expansion coefficients 
of the ground-state wave function in the infinite lattice limit 
may be approximately determined via CCM calculations. The two-body 
expansion coefficients have thereby been obtained as a function of 
lattice distance. For the triangular lattice HAF, a heuristic 
rule which fits CCM results for the coefficients corresponding 
to two-body excitations with respect to the model state was 
also obtained. Certain patterns, for the triangular lattice HAF,  
were also seen for coefficients corresponding to higher-order
excitations, although no complete heuristic rule for all Ising 
states could be inferred. For the square lattice HAF, it has once 
more been noted that {\it all} of these expansion coefficients 
obey the Marshall-Peierls sign rule for the LSUB$m$ 
and SUB2 approximations with no other assumption other 
than the model state. The convergence of the Ising-expansion 
coefficients for the LSUB$m$ series of results was also seen
for small lattice distance for both lattices. In this manner,
the LSUB$m$ results are believed to provide an accurate simulation
of the true nodal surface of these models at small lattice
separation. 

Finally, suggestions were made with respect to further 
QMC calculations for the triangular lattice HAF. The 
first was the inclusion of a `diamond-shaped'
cluster in a FNQMC calculation, which we believe to be the
most important cluster not yet used in such a calculation.
However, we also infer from our results that many three-
and four-body clusters are important in order to accurately
simulate this model. One method of including these 
correlations in a QMC calculation would be to use the 
nodal surface predicted by CCM calculations.

\section*{Acknowledgements}

One of us (RFB) gratefully acknowledges a research grant from the
Engineering and Physical Sciences Research Council (EPSRC) of Great
Britain. This work has also been supported in part by the Deutsche
Forschungsgemeinschaft (GRK 14, Graduiertenkolleg on `Classification of 
phase transitions in crystalline materials').

\pagebreak


\newpage
\begin{figure}
\epsfxsize=11cm
\centerline{\epsffile{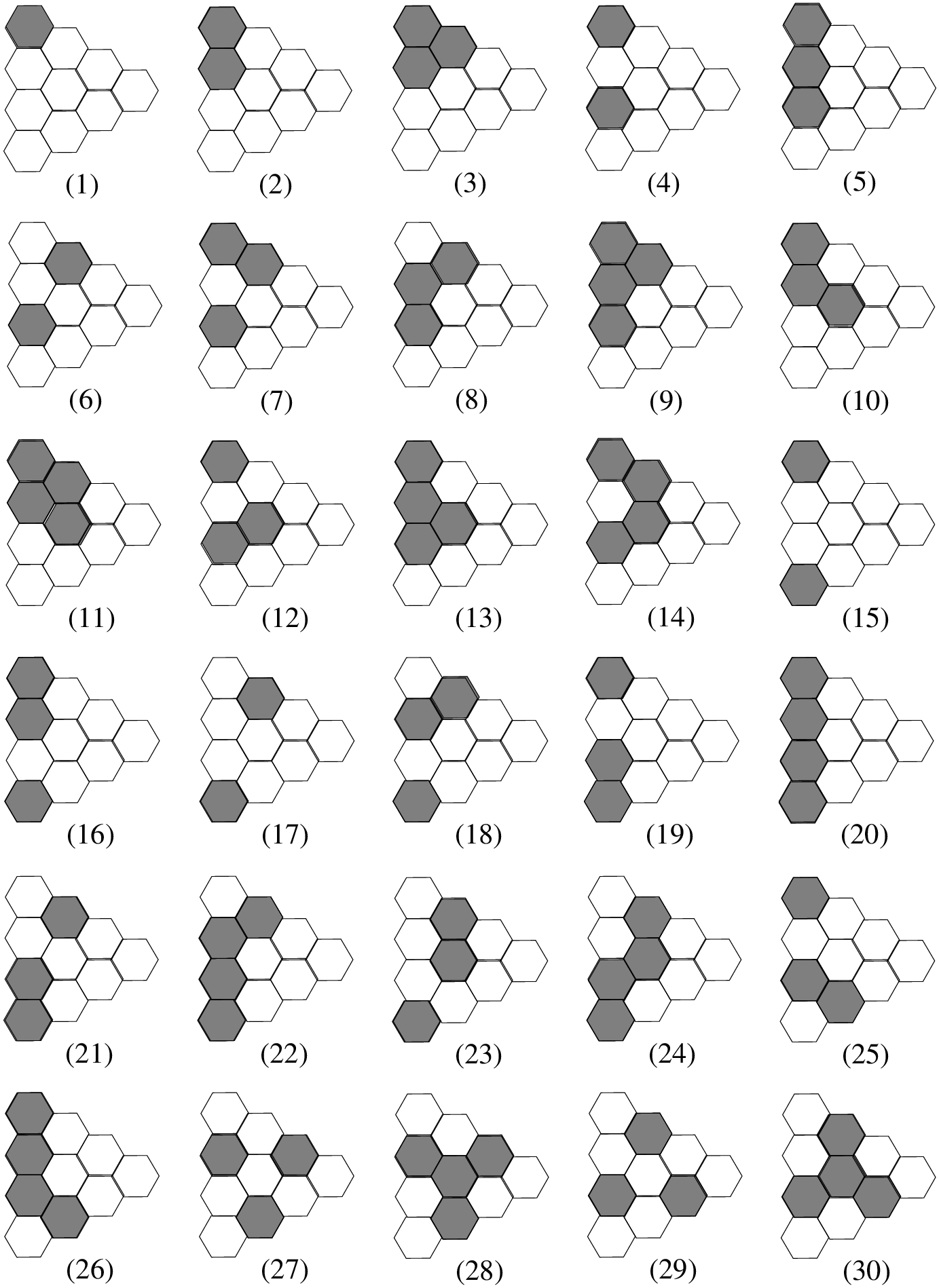}}
\makebox[1cm]{}

\makebox[1cm]{}

\makebox[1cm]{}

\makebox[1cm]{}

\makebox[1cm]{}

\caption{Excitation cluster configurations for the triangular lattice  
  antiferromagnet within the LSUB4 approximation. Each hexagon 
  marks the lattice position of a spin raising operator applied to the 
  model state at a given point on the triangular lattice.}
\label{fig3}
\end{figure}

\newpage
\begin{figure}
\epsfxsize=11cm
\centerline{\epsffile{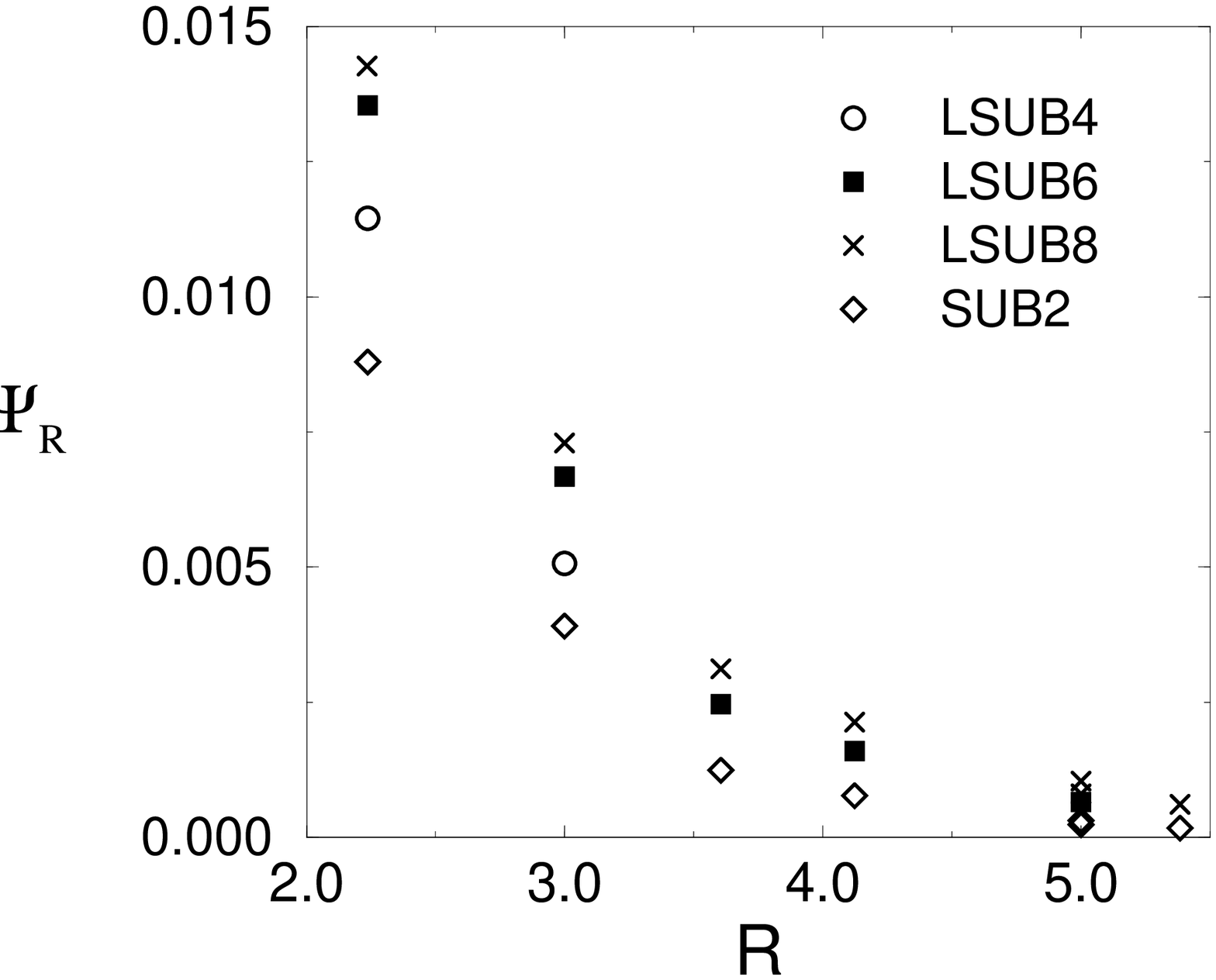}}
\makebox[1cm]{}
\caption{Results for the Ising-expansion coefficients, 
plotted as a function of the lattice distance 
$R$, corresponding to two-body excitations with respect to the 
model state for the spin-half, square lattice HAF ($x=1$)  
obtained via the LSUB$m$ approximation scheme (with 
$m=\{4,6,8\}$) and the SUB2 approximation.}
\label{fig2}
\end{figure}

\newpage
\begin{figure}
\epsfxsize=11cm
\centerline{\epsffile{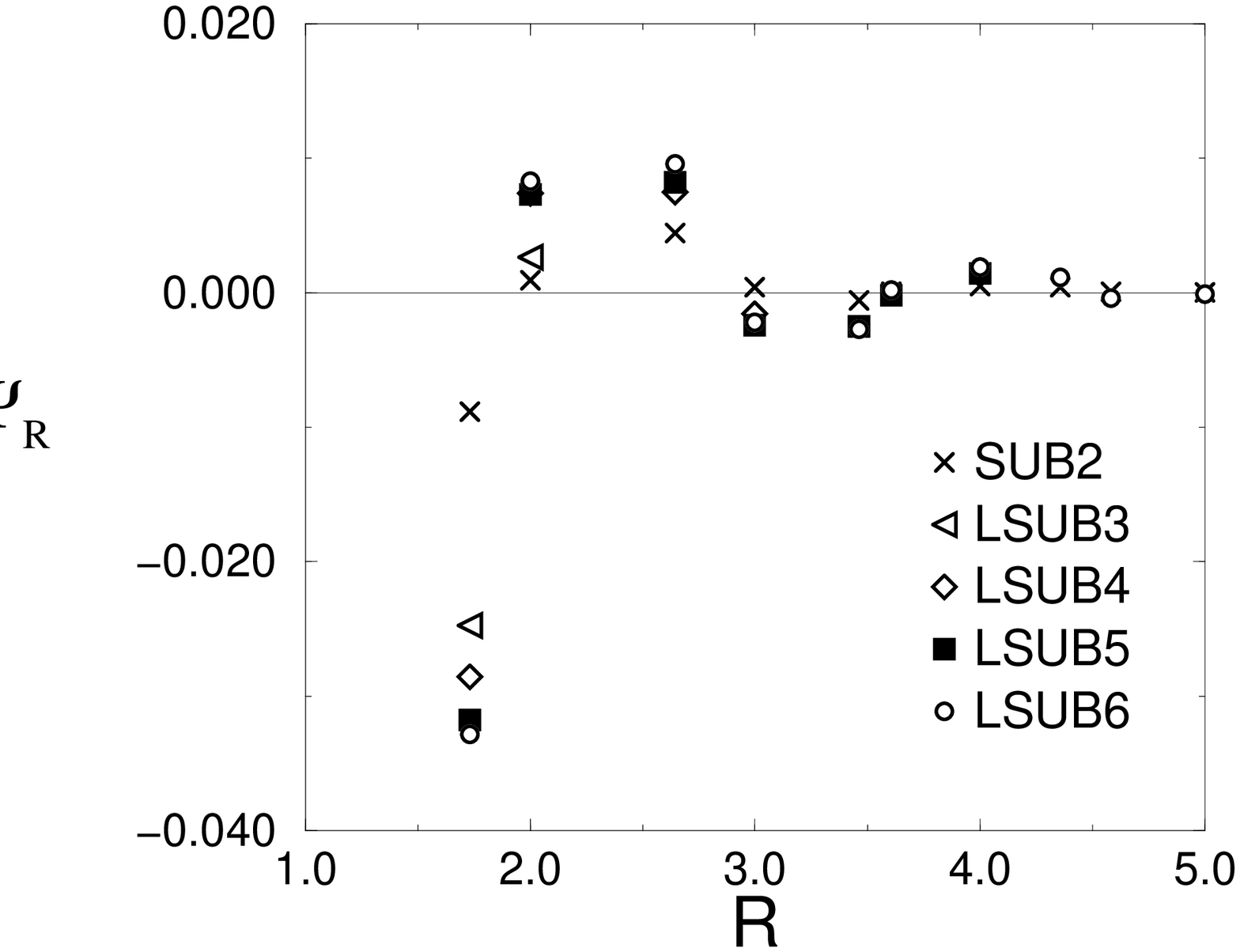}}
\makebox[1cm]{}
\caption{Results for the Ising-expansion coefficients, 
plotted as a function of the lattice distance 
$R$, corresponding to two-body excitations with respect to the model 
state for the spin-half, triangular lattice HAF ($\lambda=1$) 
obtained via the LSUB$m$ approximation scheme (with 
$m=\{3,4,5,6\}$) and the SUB2 approximation.}
\label{fig1}
\end{figure}

\newpage
\noindent
{Table 1. Expansion coefficients $\{ \Psi_I \}$ for the spin-half, 
  triangular lattice HAF ($\lambda=1$) for the clusters
  given in Fig. \ref{fig3} within LSUB$m$ (with $m=\{3,4,5,6\}$) 
  approximation scheme and the SUB2 approximation. For 
  the sake of consistency, the only clusters 
  expansion coefficients $\{ \Psi_I \}$ that have been determined 
  are those which are also used within the given approximation.}
\begin{center}
{\scriptsize
\begin{tabular}{|c|c|c|c|c|c|}   \hline
Cluster       &SUB2        &LSUB3         &LSUB4         &LSUB5      
&LSUB6       \\ \hline
1             &0.000000    &0.0000000     &0.000000      &0.000000   
&0.000000    \\ \hline
2             &0.114444    &0.1280959     &0.141570      &0.145502           
&0.149245    \\ \hline
3             &--          &0.0000000     &0.000000      &0.000000   
&0.000000    \\ \hline
4             &0.000941    &0.0026416     &0.007384      &0.007310           
&0.008302    \\ \hline
5             &--          &0.0000000     &0.000000      &0.000000   
&0.000000    \\ \hline
6             &$-$0.008838 &$-$0.024744   &$-$0.028535   &$-$0.031743      
&$-$0.032843 \\ \hline
7             &--          &--            &0.004098      &0.004380           
&0.004694    \\ \hline
8             &--          &$-$0.029600   &$-$0.039855   &$-$0.044560   
&$-$0.047332 \\ \hline
9             &--          &--            &0.020290      &0.021182           
&0.022933    \\ \hline
10            &--          &0.029600      &0.039855      &0.044560           
&0.047332    \\ \hline    
11            &--          &--            &0.055163      &0.060557           
&0.065889    \\ \hline
12            &--          &--            &$-$0.004098   &$-$0.004380          
&$-$0.004694 \\ \hline
13            &--          &--            &0.020290      &0.021182
&0.022933    \\ \hline
14            &--          &--            &0.014343      &0.013776   
&0.014106    \\ \hline
15            &0.000428    &--            &$-$0.001568   &$-$0.002414        
&$-$0.002185 \\ \hline
16            &--          &--            &0.000133      &0.000437
&0.000409    \\ \hline
17            &0.004417    &--            &0.007473      &0.008187
&0.009555    \\ \hline
18            &--          &--            &$-$0.001828   &$-$0.002663       
&$-$0.002865 \\ \hline
19            &--          &--            &$-$0.000133   &$-$0.000437         
&$-$0.000409 \\ \hline
20            &--          &--            &0.019778      &0.020721
&0.022046    \\ \hline
21            &--          &--            &0.005219      &0.006398   
&0.007213    \\ \hline
22            &--          &--            &0.025031      &0.027608           
&0.029788    \\ \hline
23            &--          &--            &$-$0.005219   &$-$0.006398   
&$-$0.007213 \\ \hline
24            &--          &--            &0.013840      &0.012139
&0.012059    \\ \hline
25            &--          &--            &0.001828      &0.002663   
&0.002865    \\ \hline
26            &--          &--            &0.025031      &0.027608           
&0.029788    \\ \hline
27            &--          &--            &0.004266      &0.006421   
&0.007184    \\ \hline
28            &--          &--            &0.000679      &0.004521           
&0.006052    \\ \hline
29            &--          &--            &$-$0.004266   &$-$0.006421       
&$-$0.007184 \\ \hline
30            &--          &--            &0.000679      &0.004521
&0.006052    \\ \hline
\end{tabular}
}
\end{center}


\begin{thebibliography}{99}

\bibitem{ref1} F. Coester,
                {\sl Nucl. Phys.} {\bf 7}, 421 (1958);  
                F. Coester and H. K\"ummel, {\em ibid.} {\bf 17}, 477 (1960).

\bibitem{ref2} J. \v{C}i\v{z}ek,
                {\sl J. Chem. Phys.} {\bf 45}, 4256 (1966);   
                {\sl Adv. Chem. Phys.} {\bf 14}, 35 (1969).

\bibitem{ref3} R.F. Bishop and K.H. L\"uhrmann,
                {\sl Phys. Rev. B} {\bf 17}, 3757 (1978). 

\bibitem{ref4} H. K\"ummel, K.H. L\"uhrmann, and J.G. Zabolitzky, 
                {\sl Phys Rep.} {\bf 36C}, 1 (1978).

\bibitem{ref5} J.S. Arponen, 
                {\sl Ann. Phys.} {\em (N.Y.)} {\bf 151}, 311 (1983).

\bibitem{ref6} R.F. Bishop and H. K\"ummel,
                {\sl Phys. Today} {\bf 40(3)}, 52 (1987).

\bibitem{ref7} J.S. Arponen, R.F. Bishop, and E. Pajanne, 
                {\sl Phys. Rev. A} {\bf 36}, 2519 (1987);
                {\em ibid.} {\bf 36}, 2539 (1987);
                in {\em Condensed Matter Theories}, edited by P. Vashishta,
                R.K. Kalia, and R.F. Bishop (Plenum, New York, 1987), 
                Vol. 2, p. 357.

\bibitem{ref8} R.J. Bartlett,
                {\sl J. Phys. Chem.} {\bf 93}, 1697 (1989).

\bibitem{ref9} R.F. Bishop, 
                {\sl Theor. Chim. Acta} {\bf 80}, 95 (1991). 



\bibitem{Roger} M. Roger and J.H. Hetherington, 
                {\sl Phys. Rev. B} {\bf 41}, 200 (1990).

\bibitem{Bishop1} R.F. Bishop, J.B. Parkinson, and Y. Xian, 
                  {\sl Phys. Rev. B} {\bf 43}, 13782 (1991);
                  {\sl Theor. Chim. Acta} {\bf 80}, 181 (1991);
                  {\sl Phys. Rev. B} {\bf 44}, 9425 (1991);
                  in {\sl Recent Progress in Many-Body Theories},
                  edited by T.L. Ainsworth, C.E. Campbell, 
                  B.E. Clements, and E. Krotscheck (Plenum, New York, 
                  1992), Vol. 3, p. 117. 

\bibitem{Bishop3} R.F. Bishop, R.G. Hale, and Y. Xian,
                  {\sl Phys. Rev. B} {\bf 46}, 880 (1992).

\bibitem{Bishop2} R.F. Bishop, R.G. Hale, and Y. Xian,
                  {\sl Phys. Rev. Lett.} {\bf 73}, 3157 (1994). 

\bibitem{Lo1}     W.H. Wong, C.F. Lo, and Y.L. Wang, 
                  {\sl Phys. Rev. B} {\bf 50}, 6126 (1994).

\bibitem{Bishop4} R.F. Bishop, J.B. Parkinson, and Y. Xian, 
                  {\sl J. Phys.: Condens. Matter} {\bf 5}, 9169 (1993).

\bibitem{Farnell1} D.J.J. Farnell and J.B. Parkinson, 
                   {\sl J. Phys.: Condens. Matter} {\bf 6}, 5521 (1994).

\bibitem{Xian1} Y. Xian, 
               {\sl J. Phys.: Condens Matter} {\bf 6}, 5965 (1994).

\bibitem{Bursill} R. Bursill, G.A. Gehring, D.J.J. Farnell, J.B. Parkinson, 
                  T. Xiang, and C. Zeng,
                  {\sl J. Phys.: Condens. Matter} {\bf 7}, 8605 (1995).
                
\bibitem{Xian}  Y. Xian,
               in {\sl Condensed Matter Theories}, edited by M. Casas,
               M. de Llano, and A. Polls (Nova Science Publ., Commack, 
                New York, 1995), 
               Vol. 10, p. 541.

\bibitem{Zeng3}  C. Zeng and R.F. Bishop,  
                 in {\sl Coherent Approaches to Fluctuations}, 
                 edited by M. Suzuki and N. Kawashima, (World 
                 Scientific, Singapore, 1996), p. 296.  

\bibitem{Bishop5} R.F. Bishop, D.J.J. Farnell, and J.B. Parkinson, 
        {\it J. Phys.: Condens.\ Matter} {\bf 8}, 11153 (1996).

\bibitem{Farnell2}  D.J.J. Farnell, S.A. Kr\"uger, and J.B. Parkinson,
        J. Phys. Condens. Matter. {\bf 9}, 7601 (1997); 

\bibitem{Zeng1} C. Zeng, D.J.J. Farnell, and R.F. Bishop, 
  {\it J. Stat. Phys.}, {\bf 90}, 327 (1998).

\bibitem{Zeng2} C. Zeng. I. Staples, and R.F. Bishop, 
  {\it J. Phys.: Condens. Matter} {\bf 7}, 9021 (1995);
  {\it Phys. Rev. B} {\bf 53}, 9168 (1996).

\bibitem{Bishop6} R.F. Bishop, D.J.J. Farnell, and J.B. Parkinson,
  {\it to be published.}



\bibitem{Marshall1} W. Marshall, 
  {\it Proc.\ R. Soc.\ London A}\ {\bf 232}, 48 (1955).



\bibitem{Richter1} J. Richter, N.B. Ivanov and K. Retzlaff, {\it Europhysics 
    Letters} {\bf 25}, 545 (1994); 
  A. Voigt, J. Richter, N. B. Ivanov, 
  {\sl Physica A} {\bf 245}, 269 (1997).




\bibitem{Anderson1} P.W. Anderson,
                   {\sl Phys. Rev.} {\bf 86}, 694 (1952);
                   T. Oguchi,
                   {\em ibid.} {\bf 117}, 117 (1960).

\bibitem{Singh2}  R.R.P. Singh,
                  {\sl Phys. Rev. B} {\bf 39}, 9760 (1989); 
                  W. Zheng, J. Oitmaa, and C.J. Hamer, 
                  {\sl ibid.} {\bf 44}, 11869 (1991).                   

\bibitem{Carlson} J. Carlson,
                  {\sl Phys. Rev. B} {\bf 40}, 846 (1989);
                  N. Trivedi and D.M. Ceperley,
                  {\sl ibid.} {\bf 41}, 4552 (1990).

\bibitem{Runge1}   K.J. Runge, 
                   {\sl Phys. Rev. B} {\bf 45}, 12292 (1992);
                   {\sl ibid.} {\bf 45}, 7229 (1992). 

\bibitem{Huse} D.A. Huse and V. Elser, 
               {\sl Phys. Rev. Lett.} {\bf 60}, 2531 (1988). 



\bibitem{Singh1} R.R.P. Singh and D.A. Huse,
  {\it Phys. Rev. Lett.} {\bf 68}, 1766 (1992).  

\bibitem{Jolicoeur1} T. Jolicoeur and J.C. LeGuillou, 
  {\it Phys. Rev. B} {\bf 40}, 2727 (1989).

\bibitem{Boninsegni1} M. Boninsegni, 
  {\it Phys. Rev. B} {\bf 52}, 5304 (1995). 

\bibitem{Bernu1}  B. Bernu, P. Lecheminant, C. Lhuillier, and L. Pierre, 
  {\it Phys. Scripta} {\bf T49}, 192 (1993);
  {\it Phys. Rev. B} {\bf 50}, 10048 (1994). 

\bibitem{Henley1}  C.L. Henley and B.E. Larson, {\it 
        J. Appl. Phys.} {\bf 67}, 5752 (1990). 

\end{thebibliography}
\end{document}